\documentstyle[12pt]{article}
\normalsize

\def\a{\alpha}
\def\b{\beta}
\def\d{\delta}
\def\k{\kappa}
\def\l{\lambda}
\def\m{\mu}

\def\p{\psi}
\def\r{\rho}

\def\pa{\partial}

\let\oldtheequation=\theequation
\def\doteqs#1{\setcounter{equation}{0}
            \def\theequation{{#1}.\oldtheequation}}

\newcounter{sxn}
\def\sx#1{\addtocounter{sxn}{1} \bigskip\medskip \goodbreak \noindent{\large\bf
\centerline{\thesxn.~~#1}} \nobreak \medskip}
\def\sxn#1{\sx{#1} \doteqs{\thesxn}}

\newcounter{axn}

\def\br{\begin{thebibliography}{abc}}
\def\er{\end{thebibliography}}

\date{}

\begin{document}
\bibliographystyle{unsrt}
\footskip 1.0cm
\thispagestyle{empty}
\begin{flushright}
UR - 1310\\
ER-40685-760\\
May 1993\\
\end{flushright}
\vspace*{10mm}
\begin{center}{\LARGE RENORMALIZATION IN\\
                QUANTUM MECHANICS\\}
\vspace*{10mm}
{\large   K. S. Gupta and
          S. G. Rajeev. \\ }
\vspace*{10mm}
{\it Department of Physics,\\
University of Rochester,\\
Rochester, NY 14627, USA}.\\
\newcommand{\bc}{\begin{center}}
\newcommand{\ec}{\end{center}}
\ec
\vspace*{5mm}

\normalsize
\centerline{\bf ABSTRACT}

\vspace*{5mm}

We implement the concept of Wilson renormalization in the context of simple
quantum mechanical systems. The attractive inverse square potential leads to a
$\b$ function with a nontrivial ultraviolet stable fixed point and the Hulthen
potential exhibits the crossover phenomenon. We also discuss the implementation
of the Wilson scheme in the broader context of one dimensional potential
problems. The possibility of an analogue of Zamolodchikov's
$C$ function in these systems is also discussed.

\newpage
\newcommand{\be}{\begin{equation}}
\newcommand{\ee}{\end{equation}}

\baselineskip=24pt
\setcounter{page}{1}
\sxn{INTRODUCTION}

	Renormalization is a familiar concept
in the context of field theory. Although the modern theory of renormalization
as formulated by Wilson \cite{wilson} is completely nonperturbative in nature,
in many actual field theory calculation such a nonperturbative scheme is
often impossible to implement in practice. It is therefore interesting to look
for situations where the concept of renormalization can be actually implemented
within a nonperturbative scheme. Quantum mechanics happens to provide us with
such a situation.

	The concept of renormalization is associated with a $\beta$
function whose zeroes correspond to the fixed points of the theory. However, in
most field theories, the $\beta$ function can only be calculated within the
perturbation theory. It would therefore be interesting to find examples where
the $\beta$ function can be calculated nonperturbatively and its zeroes can be
determined thereby. Quantum mechanics again promises to present such a
framework.

	The study of renormalization in quantum mechanics however is not
entirely new. Such a concept has been used in the literature to treat singular
power law type potentials \cite{frank,case}. More recently
Thorn \cite{thorn} and Jackiw \cite{jackiw}
considered the renormalization of a delta function potential in two
and three dimensions. In two dimensions,
this is a perfectly well posed eigenvalue problem,
but the usual analysis does not yield a bound state.
The concept of renormalization was introduced to produce a well defined
bound state in that problem.

	In this paper we carry out a similar analysis for the inverse square
potential in one dimension and the Hulthen potential.
The inverse square potential has been treated in the
literature \cite{case,parisi} in the past. Such a potential is of interest
in polymers \cite{enzo}. However, as we show in Section 2,
the usual analysis states produces a ground state with
arbitrarily negative energy,
thereby rendering the solution physically meaningless.
Therefore, in this particular case, the problem as it stands is not well
posed. This is an important conceptual
difference between the inverse square potential problem and that treated by
Thorn \cite{thorn}. In Section 3 we show how to remedy this
using the concept of renormalization. Section 4 deals with the Hulthen
potential. This
approaches the Coulomb potential for short distances and therefore has no
inherent instability. However there is an interesting renormalization of this
problem which exhibits the crossover phenomenon. We would like to stress
that renormalization is a general principle whose relevance
is independent of the existence of any singularity in the problem.
In Section 5
we sketch a method of implementing the full Wilson
renormalization scheme in a more general context of one dimensional
potential problems.
Invariance of the full spectral data under scale transformation
leads to a definite expression of the $\beta$ function. We indicate the method
for obtaining the renormalized coupling constants using the inverse scattering
method. We also speculate on the possibility of obtaining an analogue of the
$C$ function of two dimensional conformal field theory in the context of
quantum mechanics. Section 6 summarizes and concludes the paper.

\sxn{THE INVERSE SQUARE POTENTIAL}

	The Hamiltonian for the one dimensional inverse square potential is
given by
\be
H=-{d^2\over dx^2}-{\alpha\over x^2},
\ee
where a positive $\a$ correspond to an attractive potential. Eqn. 1 can be
regarded as the radial equation for an s-wave state so that $x$ has a range
$0 \leq x < \infty$. The above Hamiltonian can be decomposed as
\be
H = ({d\over dx}+{{\nu+ 1/2}\over x})(-{d\over dx}+{{\nu+ 1/2}\over x}),
\ee
where $\a = \nu(1 - \nu)$.
Therefore for $0 \leq \a \leq 1/4$, $H$ is a positive bilinear form on the
space of smooth functions of compact support vanishing at the origin. We denote
the latter by $C_0^{\infty}(R^+)$.

	Next we note that the operator
\be
H_0 = -{d^2\over dx^2}
\ee
is self-adjoint in the domain $D_0 =\{\psi \in L^2(R^+) \mid \psi(0) = c
\psi^{\prime}(0), c \in R\}$. The operator $H$ can be written as
\be
H = H_0 + H_1,
\ee
where,
\be
H_1 = -{\alpha\over x^2}.
\ee
In general, $H$ is not self-adjoint in the domain in the domain $D_0$. However,
if $H_1$ is in some sense ``small" compared to $H_0$, then $H$ is actually
self-adjoint in the domain $D_0$.
In our case, it can be shown that if $0 \leq \a
\leq 1/4$, then $H$ is actually self-adjoint in the domain $D_0$. [Technically
$H_1$ is said to be relatively form-bounded with respect to $H_0$
\cite{simon}.]

	Let us next look at the case $\a \geq 1/4$. In this range $H_1$ is not
form bounded by $H_0$. However, since
$H$ is real and symmetric in $C_0^\infty(R^+)$, by von Neumann's theorem
\cite{simon} it would have self-adjoint extensions. Ordinarily, once an
operator is self-adjoint it defines a well-posed eigenvalue problem. However in
this case, even though the eigenvalue problem is well-posed mathematically and
has been solved \cite{case}, the ground state has energy $- \infty$ and hence
is physically meaningless.
There are normalizable wavefunctions belonging to $C_0^\infty(R^+)$
of arbitrarily negative energy and no self-adjoint extension is going to
remove this problem. To see this consider the expression
\be
{(\p, H \p) \over (\p, \p)} = {\int \mid \p^{\prime}(x)\mid^2 dx
- \a \int {\mid \p (x)\mid^2 \over x^2} d x \over \int\mid \p (x)\mid^2 dx}
\ee
where $\p \in C_0^\infty(R^+)$. Let $\p_\l (x) = \surd\l \p (\l x)$. Then
\be
{(\p_\l, H \p_\l) \over (\p_\l, \p_\l)} = \l^2
{\int \mid \p^{\prime}(x)\mid^2 dx
- \a \int {\mid \p (x)\mid^2 \over x^2} d x \over \int\mid \p (x)\mid^2 dx}.
\ee
Thus if
\be
\a > {\int \mid \p^{\prime}(x)\mid^2 dx \over
\int {\mid \p (x)\mid^2 \over x^2} d x},
\ee
then $H$ would be unbounded from below. So we can ask the question that what is
the minimum value of $\a$ such that the above inequality holds. This is a
problem of minimizing $\a$ over $\p \in C_0^\infty(R^+)$. Let us choose as a
trial wavefunction
\be
\p = \surd x h_{1} (x) h_2 (x),
\ee
where
$h_1 (x)$ is $0$ below $x = a/2$, rises from $0$ to $1$ smoothly between
$a/2 \leq x \leq a$ and stays constant after that. Similarly $h_2 (x)$ is
equal to $1$ below $x = b$, decreases smoothly from $1$ to $0$ between
$b \leq x \leq 2 b$ and stays constant after that. We also assume that
$a < b$. Then it is easy to show that
\be
{\int \mid \p^{\prime}(x)\mid^2 dx \over
\int {\mid \p (x)\mid^2 \over x^2} d x} \leq
{{1\over 4}~ {\rm ln}{b\over a} +
{{\rm ln}2 \over 2} + C \over {\rm ln}{b\over a}}
\ee
for any ${b\over a}$ where $C$ is a constant independent of $a$ and $b$.
If we now let ${b\over a} \rightarrow \infty$, we get
\be
\a_c \leq {1\over 4}
\ee
where $\a_c$ is the critical value of $\a$ for which the energy becomes
negative. Eqn. (2.2) already shows that $\a_c > 1/4$. Hence we can
conclude that $\a_c = 1/4$. Therefore we have shown that for $\a > 1/4$ the
hamiltonian $H$ is unbounded from below on $C_0^\infty(R^+)$ as a quadratic
form.

\sxn{RENORMALIZATION}

	We have seen in the previous section that for $\a > 1/4$, $H$ is not
bounded from below on $C_0^\infty(R^+)$. In this section we redefine the
problem so that the ground state energy is finite.
In what follows, $\a$ will be taken to be
greater than $1/4$ and we write $\a = \m^2 + 1/4$. Next we introduce a cutoff
$a$ and modify the Hamiltonian $H$ such that
\be
H=-{d^2\over dx^2}-{\alpha (a)\over x^2},
\ee
where $\a (a)$ is now considered to be a function of $a$ and $H$ is
defined on the domain consisting of smooth functions
in the range $[a, \infty]$ satisfying the condition
\be
\p (a) = 0.
\ee
$H$ is then self-adjoint and bounded below by $-\a (a)/a^2$. We will determine
the dependence of $\a (a)$ on $a$ by requiring that the ground state energy be
independent of $a$.

	Let us now consider the eigenvalue problem of the Hamiltonian $H$ in
the bound state sector of the theory. This is given by
\be
H \p = - \k^2 \p, ~~~~~~~~\k \in R.
\ee
This can be written as
\be
{{d^2 \p}\over dz^2}-[{{\m^2 + {1 \over 4}}\over x^2} - 1]\p
= 0,~~~~z = \k a.
\ee
The above equation is solved by
\be
\p ( z) = \surd z K_{i \m}(z),
\ee
where $K_{i \m}(z)$ is the modified Bessel function of the third kind
( also called the Macdonald function ) of order $i \m$ which is purely
imaginary. For small values of $z$ and $\m$, the zeroes of this function,
denoted by $z_n$ are given by
\be
z_n = e^{({- n \pi \over \m })}~(2 e^{-\gamma)}~[1 + {\rm O}(\m)],
\ee
where $\gamma$ is the Euler's constant and $n = 1, 2, ...., \infty$.
Therefore the energy levels are given by
\be
E_n = - e^{({- 2 n \pi \over \m })}~[{2 \over {a e^{\gamma}}}]^2 [1 +
{\rm O}(\m)].
\ee
We can immediately see from the above equation that if the cutoff $a$ is
taken to zero keeping $\m $ fixed, the energy levels go to negative
infinity. However we can now regard $\m$ as a function of the cutoff $a$
and ask how $\m (a)$ must depend
on $a$ such that when the latter is taken to zero, the ground state energy
$E_1$ remains independent of $a$. This is going to fix the dependence of $\m
(a)$ on $a$ and hence give the $\beta$ function. Let us define the $\beta$
function as
\be
\beta (\m) = -a {{d \m} \over {d a}}.
\ee
The condition that the ground state energy remains independent of the cutoff as
the latter is taken to zero can be written as
\be
a {{d E_1} \over {d a}} = 0
\ee
This gives
\be
\beta (\m) = - {\m^2 \over \pi} + ....
\ee
Thus we see that our system has an ultraviolet stable fixed point at $\m = 0$
which is same as $\a = {1 \over 4}$.

	The renormalization carried out so far has used the bound state sector
of the Hamiltonian. In the usual field theoretic treatments renormalization is
performed on the scattering sector of the theory and the same analysis can be
done here too. Below we show how to perform such an analysis.

	In the scattering sector, the energy is positive and the
${\rm schr \ddot {o} dinger's}$ equation looks like
\be
H \p = k^2 \p,~~~~k \in R,~~~~\p(a) = 0.
\ee
This is the same as
\be
{{d^2 \p}\over dz^2}+[{{\m^2 (a) + {1 \over 4}}\over z^2} + 1]\p
= 0,~~~~z = k a,~~~~\p(k a) = 0.
\ee
The solution of the above equation is given by
\be
\p(z) = \surd z (A J_{i \m} (z) + B J_{- i \m} (z)),
\ee
where $A$ and $B$ are constants.
The boundary condition implies that
\be
{B \over A} = -{J_{i \m}(\k a) \over J_{- i \m}(\k a)}.
\ee
Therefore upto an overall constant, the solution is given by
\be
\p(z) = \surd z ( J_{i \m} (z) J_{- i \m}(\k a)
- J_{- i \m} (z) J_{i \m}(\k a)).
\ee
Using the large $z$ asymptotic expression for
$J_{\pm i \m}(z)$ given by
\be
J_{\pm i \m}(z) = \surd ({2 \over {\pi z}} \cos(z \mp
{\pi \over 2} i \mu - {\pi \over 4}) + ....,
\ee
the solution can be written as
\begin{eqnarray}
\p &=& \surd ({1 \over {2 \pi}})e^{i z}[J_{-i \m}(\k a) e^{\pi({\m \over 2}
- {i \over 4})} - J_{i \m}(\k a) e^{-\pi({\m \over 2}
+ {i \over 4})}] \nonumber \\
& &~+e^{- i z}[J_{-i \m}(\k a) e^{\pi({-\m \over 2}
+ {i \over 4})} - J_{i \m}(\k a) e^{\pi({\m \over 2}+ {i \over 4})}] + ...
\end{eqnarray}
Let us denote the phase shift by $\delta$. Using the fact that as
$\m \rightarrow 0,~~J_{i \m}(\k a) \rightarrow (\k a)^{i \m}$ we get
\be
e^{2 i (\d + {\pi \over 4})} =
{{e^{\m (i \log (\k a) - {\pi \over 2})}-e^{\m (-i \log (\k a) +
{\pi \over 2})}} \over
{e^{\m (i \log (\k a) + {\pi \over 2})}-e^{\m (-i \log (\k a) -
{\pi \over 2})}}} + ...
\ee
Let $ \delta^{\prime} = - \d + {\pi \over 4}$. Then from the above expression
we get
\be
\tan (\d^{\prime}(\k)) = \tan(\m \log (\k a)) \coth ( \m {\pi \over 2}) + ...
\ee
Eqn. (3.19) gives the phase shift at an arbitrary momentum $\k$.
However the phase shift given by the above equation blows up as  the cutoff is
taken to zero. We can again regard $\m$ as a function of $a$ and
ask the question how $\m$ must depend on $a$ such that the
phase shift at any particular momentum $\k_0$ be independent of $a$ as the
latter is taken to zero. This fixes the relation between $\m (a)$ and $a$,
which is now given by
\be
\m (a) = { {- n \pi } \over {\log \k_0 a - \pi {C \over 2}}},
\ee
where $C$ is a constant given by $C = (\tan \d ^{\prime}) \mid_{\k = \k_0}$
and $n = 1, 2, 3...$. From this we get the $\beta$ function as
\be
\beta = - {\m^2 \over {n \pi}} + ...
\ee
For $n = 1$, we recover the result given by eqn. (3.10).

	We have performed the renormalization by keeping the phase shift at a
particular value of $\k = \k_1^2$. Using eqn. (3.19), we can now
compute the phase shift at any other arbitrary value of $\k$ as the cutoff $a$
is taken to zero. This is given by
\be
\tan (\d^{\prime}(\k)) = \tan (\d^{\prime}(\k_1)) + {2 \over \pi}
\log {\k \over \k_1}.
\ee

	The $ S $ matrix of the theory is given by $S = e^{2 i \d} $.
Using eqns.  (3.18) and (3.19) this can be written as
\be
S = - e^{-i {\pi \over 2}}{{1 - i \tan(\d^{\prime}(\k))} \over
{1 + i \tan(\d^{\prime}(\k))}}.
\ee
The bound states can be identified with the poles of the $S$ matrix, which
has a pole when $\tan(\d^{\prime}(\k) = i$. If the corresponding $\k$ be
denoted by $\k_{{\rm bound}}$ then the latter is given by
\be
\k_{{\rm bound}} = \k_1~e^{- {\pi \over 2} \tan(\d^{\prime}(k_1))}.
\ee
Therefore it seems that the $S$ matrix has only one pole and correspondingly
only one bound state. However, we also note that the $S$ matrix has a branch
point at $\k = 0$. Thus $\k = 0$ is an accumulation point of poles of the $S$
matrix.

	We note in the passing that for $\a \leq 1/4$ the scattering problem is
perfectly well defined. If we write $\a = 1/4 - \nu^2$ then the phase shift
in this case is given by
$\d = - {\pi \over 2} ( \nu + {1 \over 2})$, independent of the momentum. This
is a consequence of scale invariance. Scale invariance is broken by
renormalization for $\a > {1 \over 4}$. As $x \rightarrow \infty$,
$\a \rightarrow {1 \over 4}$ and $\d \rightarrow - {\pi \over 4}$ and the two
results match.

\sxn{THE HULTHEN POTENTIAL}

	The Hulthen potential is defined by
\be
V(r) = - {\a \over a^2} {e^{-{r \over a}} \over {1 - e^{-{r \over a}}}},
\ee
where $\a$ and $a$ are constants. For $r$ small compared to $a$, the Hulthen
potential approaches the Coulomb potential whereas for $r$ large compared to
$a$ it approaches zero exponentially fast. Therefore $a$ can be thought of as
an infrared regulator for the Coulomb problem. The bound states of this problem
has energy levels given by \cite{flugge}
\be
E_n = - {1 \over a^2} { {(\a - n^2)^2} \over {4 n^2}}.
\ee

	Proceeding along the lines of the earlier section we now regard $\a$ as
a function of $a$ and ask how $\a (a)$ must depend on $a$ so that the
ground state energy is independent of $a$. This defines the $\beta$ function
which is then given by
\be
\beta (\a) = \a - 1.
\ee
Therefore the $ \beta $ function has a zero at $\a = 1$ and that is an infrared
stable fixed point.

	In order to understand the crossover phenomenon, let us ask as to how
the $\beta$ function behaves as $\a \rightarrow \infty$.
Recall that the $\beta$ function is really a vector
field, so that  under a change of variables $\a \rightarrow w$,
$\b (\a){\pa \over {\pa \a}} = \tilde {\b}(w){\pa \over {\pa w}}$
where $\tilde {\b} (w)$ is the component of the $\b$ vector field in the new
variables. It is given by
\be
\tilde{\b}(w) = -w ( 1 - w).
\ee
As $\a \rightarrow \infty$, $w \rightarrow 0$ from above and therefore in that
limit $1 -w$ is positive. Therefore $w = 0$ is an
ultraviolet stable fixed point. In terms of the old coordinates this fixed
point appears at an infinite value of $\a$. Thus we see that this system has
two fixed points one of which is infrared stable and the other one is
ultraviolet stable. Therefore, a system described by the Hulthen potential
exhibits the crossover phenomenon.

\sxn{GENERAL DISCUSSION OF THE $\b$ FUNCTION}

	Until this point we have shown a couple of examples where the process
of renormalization was performed in quantum mechanical systems by demanding the
invariance of the phase shift or the ground state energy under the
renormalization group transformations. In this section we shall sketch a
general method of implementing the renormalization group transformations in the
space of one dimensional potential problems.
The basic strategy for renormalization would be the same
as displayed in the previous sections : given a potential we will first
introduce a cutoff so that the solution to the
${\rm Schr \ddot {o}dinger's}$ equation depends on
the cutoff. Here we consider the entire
spectral data of the problem instead of just the phase shift or the ground
state energy. The potential in general would be a function of many coupling
constants and the corresponding spectral data would depend on the momentum, the
cutoff and all the coupling constants. We can then ask as to how the potential
or the coupling constants must depend on the cutoff so that the spectral data
is independent of the cutoff as the latter is removed. This will fix the
$\beta$ function of the problem.

	Let the system be described by the ${\rm Schr \ddot {o}dinger's} $
equation
\be
-{d^2\over dr^2}\psi + V(r) \psi = k^2 \psi = \l \psi,
\ee
where $k^2 = \l$ is the energy. Let $a$ denote the cutoff. Then the boundary
conditions are given by $\p (k, a) = 0$ and $\p^{\prime}(k, a) = 1$.
We shall also assume ( for technical reasons ) that
\be
\int_a^\infty r^j \mid V(r) \mid dr < \infty,~~~j=1,2.
\ee

	A system described by (5.1) - (5.2)
would in general have both bound states
and scattering states. In the scattering sector $k$ is real and
$ > 0$. Let $f(k, r)$ be one of the two linearly independent solutions of (5.1)
which in the limit of $r \rightarrow \infty$ satisfies the condition
$e^{i k r} f(k, r) \rightarrow 1$. Let $f(k, a) $ be denoted by $f(k)$.
Then it can be shown that as $r \rightarrow \infty$,
\be
\psi(k, r) \rightarrow {{\mid f(k) \mid} \over k} \sin(k(r-a) + \eta(k)),
\ee
where $\eta(k)$ is the phase shift \cite{lev}. The bound state sector for the
above problem will be described by a finite number ( say $n$ ) \cite{lev}
of bound states
with normalization factors $c_i, i = 1,...., n$. With this information we can
now define the spectral data. Consider a monotonically nondecreasing
function $\r (\l)$, which for $\l < 0$ is constant except for jumps at energies
corresponding to those of the bound states, the magnitude of each jump being
equal to the normalization constant for that bound state. For $\l = 0$ let $\r
(\l) = 0$ and for $\l > 0$ let
\be
\r (\l) = {1 \over \pi} \int_0^{\l} {{\sqrt \l} \over {\mid f(\sqrt \l)
\mid^2}} d \l.
\ee
This function $\r (\l)$ defines the spectral data \cite{lev}.
The spectral data has a dimension of $\l^
{3 \over 2}$ \cite{lev} and we shall therefore divide it by $\l^ {3 \over 2}$
and work with this dimensionless quantity denoted by
$\tilde{\r}$. Let the cutoff be denoted by $a$. Therefore
$\tilde{\r}$ is a function
of $a$, $V(r)$ and $\l$. Since this quantity is dimensionless, it is invariant
under the simultaneous scaling of $a$, $V(r)$ and $\l$. This implies that
\be
-2 \l {{\pa \tilde {\r}}  \over {\pa \l}} + a  {{\pa \tilde {\r} }
\over {\pa a}}
+\int {{\d \tilde {\r} } \over {\d V(r)}} (2 V(r) + r{{\pa V(r)} \over
{\pa r}})dr= 0.
\ee
We must also impose the condition that the spectral data be invariant under the
renormalization group transformations. This leads to the equation
\be
a  {{\pa \tilde {\r} } \over {\pa a}}+ \int {{\d \tilde {\r} } \over
{\d V(r)}}
a {{\pa V } \over {\pa a}}dr = 0.
\ee
Let us define the $\b$ function as
\be
\b (r) = -a{{\pa V(r) } \over {\pa a}}+(2 V(r) + r{{\pa V(r)} \over {\pa r}})
\ee
The above equations then imply that
\be
-2 \l {{\pa \tilde {\r} } \over {\pa \l}}+ \int {{\d \tilde {\r} }
\over {\d V(r)}}\b dr= 0.
\ee
{}From this we get the $\b$ function as
\be
\b (r) = \int 2\l {{\pa \tilde {\r} } \over {\pa \l}} {{\d V(r) }
\over {\d \tilde {\r}}}d \l.
\ee
Under an infinitesimal variation of the spectral data $\r$, the change in the
potential $V(r)$ is given by \cite{lev}
\be
\d V(r) = -4 \int_{- \infty}^{\infty} \p(\l,r) \p^{\prime}(\l,r)d(\d \r (\l)).
\ee
Integrating this by parts and dropping the boundary term ( this is justified
since at $\l = \pm \infty$, $\d \r ( \l)=0$ ) we get
\be
{{\d V(r)} \over {\d \tilde{\r} (\l)}} = 4 \l^{3 \over 2}
\p (\l,r) \p^{\prime} (\l,r).
\ee
Therefore the final form for the $\b$ function is given by
\be
\b (r) = \int 8 \l^{5 \over 2}{{\pa \tilde {\r} } \over {\pa \l}}
\p (\l,r) \p^{\prime} (\l,r) d \l.
\ee

	We end this section by making a few comments about the above set of
formulae. The equation (5.7) can be written as
\be
\b (r) = -a{{d V(r,a) } \over {d a}} = -{{dV(r, t)} \over {dt}},~~~t=\log(a).
\ee
The left hand side of this equation is known from (5.12). However, integrating
such an equation to obtain the renormalized potential is an enormously
complicated task. We therefore approach this problem from a different point of
view. Let us first note that what we are calling the $\b$ function is really a
vector field. The renormalization group transformations move along the integral
curves of this vector field. Imagine now a ``theory space" described in terms
of all one
dimensional potentials or correspondingly the
collection of the spectral data (or $\tilde {\r}$.)
It is possible to
express the components of the $\b$ vector field in terms of either of these
variables which can serve as coordinates for the ``theory space". Usually the
renormalization group transformation in described in terms of the potentials or
coupling constants. However, the components of the $\b$ vector field in that
coordinate system are very complicated objects. Therefore it is hard to
integrate the $\b$ vector field in this coordinate system to obtain the
renormalized potential.
However, the renormalization
group transformation in terms of the $\tilde {\r (\l)}$ variable
is very simple and is given by
\be
\tilde {\r} (\l) \rightarrow \tilde {\r} (\l e^{- 2 t}).
\ee
One can therefore
start with a potential $V(r)$ and compute the spectral data
$\r (\l, a, V(r))$ and the corresponding $\tilde {\r} (\l, a, V(r))$.
The renormalization group scaling transformation is then applied to this
$\tilde {\r}$ to obtain a transformed $\tilde {\r}$. The potential for this
renormalized $\tilde {\r}$ can then be computed using the inverse scattering
method. This procedure would be equivalent to integrating the $\beta$ vector
field in expressed in terms of the potentials and would therefore
provide us with the renormalized potential. We would like to point out in the
passing that the renormalization group transformation on $\tilde {\r} (\l)$ is
qualitatively different from the isospectral deformations used in the
integrable models since the spectrum gets modified due to the renormalization
group transformations.

	In the examples discussed before, we have seen that
the $\b$ functions can vanish at certain values of the coupling constants
thereby leading to interesting fixed points.
The $\b$ function given by (5.12) would in general have many different
fixed points. It has
been conjectured in the context of two dimensional conformal field theories
that there exists a function, usually called the $C$ function, which is
monotonic along the renormalization group flow and assumes specific values at
the fixed points of the $\b$ function \cite{cfunction}. The $\b$ function that
we compute is given in terms of the quantity $\tilde {\r}$ which
is the usual spectral data $\r$ divided by $\l^{3 \over 2}$. The
usual spectral data is a monotonic nondecreasing function of $\l$ \cite{lev}
It is perhaps possible to construct in these systems
an explicit expression of the $C$ function in terms of $\r$
which would also exhibit monotonicity along the
renormalization group flow. Such a function would be of great
interest in the conceptual understanding of quantum mechanics and is currently
under study.

\newpage

\sxn{CONCLUSION}

	We have studied the effect of renormalization on various quantum
mechanical systems. It has been shown that renormalization can be used to make
the bound state problem for the attractive inverse square problem physically
well defined. The $\b$ function in that cases has an ultraviolet stable fixed
point. Next we studied the Hulthen potential which is an infrared regulated
version of the Coulomb potential. Although there is no inherent instability in
this problem, the process of renormalization in this case gives a $\b$ function
which has two fixed points, one being infrared stable and the other is
ultraviolet stable. Thus we see an interesting example of the crossover
phenomenon in quantum mechanics. This example also shows that renormalization
is a completely general principle which is not necessarily related to the
existence of instability or divergence in the problem.

	We have also considered the general problem of renormalization in
a broader context of one dimensional potential problems.
A general formula for
the $\b$ function was derived which can be evaluated once the spectral data of
the corresponding direct scattering problem is known. We also speculate on the
possibility of constructing the analogue of the $C$ function of two dimensional
conformal field theory. The construction of such a function which would have a
monotonic gradient flow along the renormalization group transformations is
currently under investigation.

\bigskip

{\bf ACKNOWLEDGEMENTS}

	This work was supported in part by the US Department of Energy, Grant
No. DE-FG02-91ER40685.

\end{document}